\documentclass[showpacs,preprintnumbers,amsmath,amssymb,twocolumn]{revtex4}

\usepackage{graphicx}
\usepackage{dcolumn}
\usepackage{bm}

\newcommand{\pr}{Phys. Rev. }
\newcommand{\aap}{Astron. Astrophys. }
\newcommand{\mnras}{Mon. Not. Roy. Astron. Soc. }
\newcommand{\nup}{Nuc. Phys. }
\newcommand{\araa}{Ann. Rev. Astron. Astrophys. }

\newcommand{\aj}{Astron. J. }
\newcommand{\plb}{Phys. Lett. }
\newcommand{\Apj}{Astrophys. J. }

\begin{document}

\title{Multiple Main Sequence of Globular Clusters\\ as a Result of
 Inhomogeneous Big Bang Nucleosynthesis}

\author{Takashi Moriya\footnote{takashi.moriya@ipmu.jp}}
\affiliation{Department of Astronomy, Graduate School of Science, University of
Tokyo, Bunkyo-ku, Tokyo 113-0033, Japan.}
\affiliation{Institute for the Physics and Mathematics of the Universe,
 University of Tokyo, Kashiwanoha 5-1-5, Kashiwa, Chiba 277-8568, Japan.}

\author{Toshikazu Shigeyama}
\affiliation{Research Center for the Early Universe, Graduate School of
Science, University of Tokyo, Bunkyo-ku, Tokyo 113-0033, Japan.}


\begin{abstract}
A new mechanism for enhancing the helium abundance in the blue main
sequence stars of $\omega$ Centauri and NGC 2808 is investigated. We suggest that
helium enhancement was caused by the inhomogeneous big bang
nucleosynthesis. 
Regions with extremely high baryon-to-photon ratios are assumed to be
 caused by the  baryogenesis. Its mass scale is also assumed to be
 $10^6M_\odot$. An example of the mechanisms to realize these two things
 was already proposed as the Affleck-Dine baryogenesis. As the baryon-to-photon ratio  becomes larger, the
primordial helium abundance is enhanced.
We calculated the big bang nucleosynthesis
and found that there exists a parameter region yielding enough helium to account for the split of the main sequence in the aforementioned globular clusters while keeping
the abundances of other elements compatible with observations.
Our mechanism predicts that heavy elements with the mass number of around 100
 is enhanced in the blue main sequence stars.
We estimate the timescales of diffusion of the enhanced helium and mass
 accretion in several stages after the nucleosynthesis to investigate
 whether these processes diminish the enhancement of helium. We found
 that the diffusion does not influence the helium content. A cloud with
 a sufficiently large baryon-to-photon ratio to account for the multiple
 main sequence collapsed immediately after the
 recombination. Subsequently, the cloud accreted the ambient matter with
 the normal helium content. If the star formation occurred both in the
 collapsed core and the accreted envelope, then the resultant star
 cluster has a double main sequence. 
\end{abstract}

\maketitle

\section{INTRODUCTION}
A globular cluster (GC) is a gravitationally bound system of stars with a
spherical shape. The
typical radius is of the order of 10 pc and its mass is about $10^{4-6}M_\odot$. Stars in a GC are known to be very
old and their chemical compositions roughly represent that of the primordial universe. 

Among known $\sim150$ GCs, $\omega$ Centauri is notorious for its unusual
properties. Its mass is about $3\times10^6M_\odot$~\cite{massofomega} and it is the
most massive GC in the Milky Way galaxy. One of the most astonishing facts about
$\omega$ Centauri is that its dwarf stars split into two
sequences~\cite{omega1,omega2,omega3}, the blue main sequence (bMS) and the red main sequence
(rMS). 25\%-35\% of the dwarf stars are populated by the bMS~\cite{omega2}. Although enhanced metallicities might split
a main sequence by lowering the surface temperature of metal-rich stars, it is reported
that stars in the bMS are more metal-rich than those in the rMS in $\omega$
Centauri~\cite{abn2}. Recently, it is reported that there also exists the third redder MS \cite{thirdMS}. This MS is suggested to be populated by super metal-rich stars 
\cite{cmd} and seems to have nothing to do with their helium contents. Furthermore, observations by the Hubble Space Telescope discovered the triple main sequence in another GC, NGC 2808~\cite{triple}. NGC 2808 is also massive, with its mass above $10^{6}M_\odot$~\cite{mass28}. It was found that, in either GCs, the
isochrone that matches the properties of the bMS can be produced by assuming
very high helium abundances ($Y$),
$0.35<Y<0.45$~\cite{abn1,abn2}. As other parameters do not have much effect on colors,
it is very likely that the helium enrichment is the cause of the multiple main
sequence.

The origin of the enhanced helium abundances in the bMS stars has
been a mystery. It was suggested, for example, that the ejecta descended from
massive asymptotic giant branch (AGB) stars would enhance the helium abundances of
the main sequence stars and split the main sequence~\cite{danto}. However, it is difficult
for massive AGB stars to produce as much helium as
$Y\sim0.35$~\cite{tsujimoto}. What is more, the amount of helium
supplied from AGB stars is not enough to increase the helium content of all
stars in the bMS up to
$\sim0.35$~\cite{tsujimoto}.
Helium enhancement by massive stars is also suggested to explain the MS
splitting \cite{supernova}. Rotating low-metal massive stars can enhance
the helium abundance in the stellar winds but such stars should accelerate the helium-rich matter by the subsequent supernova explosions.
The helium rich ejecta might be too energetic to be trapped in a
GC and could easily escape from it.
These difficulties associated with the AGB and massive star scenarios might be relieved if the GC was originally the nucleus of a more massive dwarf galaxy \cite{others}. There were much more AGB stars in the dwarf galaxy and the helium supplied from these AGB stars might be sufficient to produce the observed number of bMS stars. As for the massive star scenario, a massive dwarf galaxy might be able to trap the helium-rich ejecta due to its deeper gravitational potential.  However, there remain problems in both scenarios. It is still difficult to produce the observed fraction of bMS stars with the helium-rich matter or these scenarios seem to require an initial mass function of stars extremely biased toward massive stars. In addition, no significant spread in metallicities of stars in NGC 2808 is incompatible with the observed spread in metallicities of stars in a dwarf galaxy. 

Most helium in the universe is believed to be produced by the big bang nucleosynthesis
(BBN). By observations of extragalactic metal poor H II regions~\cite{y1}, the mass fraction of helium produced
during BBN ($Y_\mathrm{p}$) is estimated to be  $Y_\mathrm{p}\sim0.25$. One of the parameters that affect $Y_\mathrm{p}$ is the baryon-to-photon ratio ($\eta$). Recently,
results of
the five-year observation of the cosmic microwave background radiation
(CMBR) by the Wilkinson Microwave Anisotropy Probe (WMAP) were
released~\cite{wmap}. The baryon density parameter is determined to be
$\Omega_bh^2=0.02273\pm0.00062$ where $h$ is the normalized Hubble constant given by dividing the Hubble constant by $100$ $\mathrm{km\ s^{-1} Mpc^{-1}}$. Combining this value
with the well-known temperature of CMBR, $2.725\pm0.002$ K~\cite{mather}, $\eta$ is determined to be
$\eta\sim6\times10^{-10}$. Given this $\eta$, the calculation of the standard
BBN yields $Y_\mathrm{p}\simeq0.25$, matching well with the observed values~\cite{yp}.
In the standard
BBN theory, however, it is assumed that the BBN occurred homogeneously and
$\eta$ has a uniform value. Although WMAP did not detect the
inhomogeneity of the universe with scales smaller than a resolution
$l_{\mathrm{max}}\sim2000$ ($\theta_\mathrm{min}\sim0.1^\circ$), there remains
a possibility of inhomogeneity with scales under $\sim10$ Mpc. Within regions
smaller than 10 Mpc, $\eta$ might not take the standard value. Many calculations have been done
with various values of $\eta$ (see the references of~\cite{kaji}). The general results are that the higher $\eta$
becomes, the more abundant helium will be. So, here comes a new candidate
for the enigmatic mechanism of the helium enhancement in the bMS, inhomogeneous BBN with a high value of $\eta$.

We consider a baryogenesis mechanism that enhances the baryon density in regions with the mass scale of $10^6 M_\odot$, a comparable amount of baryons to the GCs that
have  multiple main sequence. Though we do not specify this baryogenesis
in more detail, the Affleck-Dine mechanism \cite{a-d} can realize this situation right after the inflation era.
This mechanism can enhance the baryon density as high as  $\eta\sim 1$
in a region with the mass scale of $10^6M_\odot$~\cite{Dol}.
The size of inhomogeneity is a free parameter in this mechanism. The mass scale and desired $\eta$ indicate that the scale of necessary inhomogeneities is much smaller than the scales of fluctuations in the CMBR that can be detected by WMAP. 
Recent papers by
Matsuura et al.~\cite{matsu1,matsu2,matsu3} have preformed calculations of the BBN with high values of $\eta$ on the basis of the Affleck-Dine baryogenesis.
Such an inhomogeneous BBN may be affected by the diffusion of neutrons. Lara \cite{lara} showed that regions with enhanced $\eta$'s do not lead to abundances of light elements significantly deviated from those of the ambient medium if their size at the temperature of 100 GK was smaller than $10^4$ cm. Since the size of the region we are concerned with in this paper is of the order of $10^{10}$ cm at this temperature (see Eq. (\ref{size})) and much larger than $10^4$ cm, the diffusion of neutrons cannot change the yields of the BBN. Still larger $\eta$'s in our scenario suppress the diffusion of neutrons further. In addition, this region extending over the event horizon even at the end of the BBN tolerates the diffusion of neutrons. 

In this paper, we seek a possibility that the over abundance of
helium in the bMS of GCs is due to the inhomogeneous BBN caused by the
 baryogenesis. First of all, we determine the parameter
region that reproduces the helium abundance of bMS stars. The parameters we concerns are the
baryon-to-photon ratio ($\eta$) and the degeneracy of electron neutrinos
($\xi_e$). $\xi_e$ is defined as $\xi_e={\mu_{\nu_e}}/{k_{\rm B}T}$ where
$\mu_{\nu_e}$ is the chemical potential of electron neutrinos and $k_\mathrm{B}$ is the Boltzmann constant. These parameters are the most effective in the sense that they can change the
abundances of elements synthesized by the BBN.
In Section II, we briefly summarize effects of the parameters $\eta$
and $\xi_e$ during the BBN. We also summarize calculations we have
done in Section II. Results are given in Section III and the evolution of the helium enhanced region is discussed in Section IV. The conclusion is given
in Section V.

\section{BIG BANG NUCLEOSYNTHESIS}
\subsection{Free Parameters}
The BBN occurred when the temperature of the universe was about
$10^{11}-10^{8}$ K. The parameters that govern the abundance of helium synthesized 
during BBN are $\eta$ and $\xi_e$ for a given set of the cosmological
parameters.
As $\eta$ represents the density of baryons, collisions of baryons occur more frequently with
increasing $\eta$, resulting in larger $Y_\mathrm{p}$. 
The parameter $\xi_e$ represents 
lepton asymmetry and it can affect $Y_\mathrm{p}$
directly
through the reactions
\begin{equation}
\mathrm{p}+\mathrm{e^-}\longleftrightarrow n+\nu_\mathrm{e},
\end{equation}
which determines the number of neutrons and protons before BBN. At the equilibrium of these reactions, the number ratio of neutrons to protons is expressed by a function of the temperature $T$ and $\xi_e$ as
\begin{equation}
\frac{n_\mathrm{n}}{n_\mathrm{p}}=\exp\left(-\frac{\Delta mc^2}{k_\mathrm{B}T}-\xi_e\right),
\end{equation}
where $\Delta mc^2$ is the difference of the rest energies between a proton and
a neutron. As most of neutrons that had existed before the BBN era were brought into
helium nuclei, a negative $\xi_e$ enhances helium synthesis.
We treat $\xi_e$ as a free parameter restricted from
observations~\cite{xi} and assume $\beta$-equilibrium at the beginning of calculations. Allowed range with 2$\sigma$ CL is
\begin{equation}
-0.01\leq\xi_e\leq0.1.
\end{equation}
Note that the Affleck-Dine mechanism prefers negative $\xi_e$ with the absolute value of the order of $\eta$ because $\eta$ is supposed to be positive \cite{kamada}. 

\subsection{Numerical Calculations}
Because the horizon scale right after the BBN is much smaller than $10^6M_\odot$ in
mass, the effect of diffusion is negligible. Thus, we calculate the BBN
with high $\eta$ in the same way as the homogeneous BBN ignoring inhomogeneities inside the region and adopt the same
method of numerical calculations as that of the ordinary homogeneous
BBN. We first calculated BBN with the reaction network including 61 nuclei
shown in Table. \ref{small}.
The nucleosynthesis with a larger reaction network with nuclei in
Table. \ref{large} is also calculated to see how much heavy elements are 
produced.
Heavy element production during BBN with high $\eta$ is discussed in \cite{matsu1,matsu2,matsu3}.
To calculate BBN, we used the Kawano code introduced
in~\cite{bbncode}, with revised reaction rates taking from~\cite{reactionrates} whenever
they are available. Reaction rates of heavy elements are taken from~\cite{hr}.

\begin{table}[t]

\begin{tabular}{cc|cc|cc}
\hline
$Z$ & $A$  & $Z$ & $A$ & $Z$ & $A$\\
\hline
n&1& C &11-15 &Mg &24-27 \\
H&1-3 &N &12-16 &Al &25-28\\
He&3,4 &O &14-19 &Si &28-31\\
Li&6-8 &F &17-20 &P &29-32 \\
Be&7,8 &Ne &19-23 &S &32 \\
B&8,10-12&Na &21-24 &&\\
\hline
\end{tabular}
\caption{Nuclei included in the small reaction network. $Z$ is the name
 of nuclei and $A$ is the mass number of $Z$ included in the network. 'n'
 refers to neutron.}
\label{small}
\end{table}

\begin{table}[htbp]

\begin{tabular}{cc|cc|cc}
\hline
$Z$ & $A$  & $Z$ & $A$ & $Z$ &$A$\\
\hline
n& 1 &K &29-70 &Sr&68-131 \\
H & 1-3 & Ca&30-73 &  Y &70-134 \\
He & 3,4,6  &Sc &32-76 &Zr&72-137 \\
Li & 6-9& Ti&34-80 & Nb&74-140   \\
Be&7-12 &V & 36-83&Mo &77-144 \\
B&8,10-14 &Cr &38-86 &Tc &79-147 \\
C&9-18 &Mn &40-89 & Ru&81-150 \\
N&11-21 &Fe &42-92 & Rh&83-153 \\
O&13-22 &Co &44-96 & Pd &86-156 \\
F&14,16-26 &Ni &46-99 &Ag &88-160 \\
Ne&15-41 &Cu &48-102 & Cd&90-163 \\
Na&17-44 &Zn &51-105 & In&92-166 \\
Mg&19-47 &Ga &53-108 & Sn&94-169 \\
Al&21-51 &Ge &55-112 &  Sb&97-172 \\
Si&22-54 &As &57-115 &Te&99-176 \\
P&23-57 &Se &59-118 & I &101-179 \\
S&24-60 &Br &61-121 & Xe  &103-182 \\
Cl&26-63 &Kr &63-124 &   & \\
Ar&27-67 &Rb &66-128 &   & \\
\hline
\end{tabular}
\caption{Nuclei included in the large reaction network. 
See the caption of Table. \ref{small} for details.}
\label{large}
\end{table}

\section{RESULTS}
We calculate the BBN with various parameters.
The calculations are started from the time
when the temperature is
$10^{11}$ K and we obtained the abundances at $10^{7}$ K. The nuclear
statistical equilibrium and the $\beta$-equilibrium are assumed at the
initial state.
\subsection{Lower Limits}
To locate a star on the bMS in $\omega$ Centauri by enhancing the helium content, the helium fraction should be as high as 0.35 \cite{omega2}.
Therefore we require $\eta$ to be greater than $\eta_{\rm b}$ that reproduces $Y_p=0.35$.
To estimate the amount of helium produced during BBN,
only the reactions of light elements are essential.
For this reason, we use the small reaction network to estimate $Y_p$.

\begin{figure}[t]
\begin{center}
\includegraphics[height=8cm,width=9cm,clip]{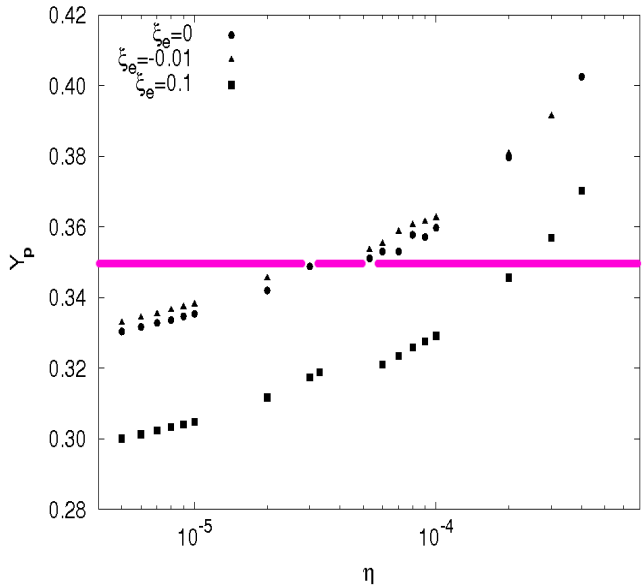}
\caption{Results of $Y_\mathrm{p}$ for some sets of $\eta$ and $\xi_e$. The horizontal
 line at the middle represents $Y_\mathrm{p}=0.35$.}
\label{600}
\end{center}
\end{figure}

The results of calculations are shown in
Fig.~\ref{600}.
We performed the calculation with three different $\xi_e$; $\xi_e=0$,
$0.01$, and $-0.1$.
The horizontal line indicates $Y_\mathrm{p}=0.35$.
For $\xi_e=0$, $\eta$ yielding $Y_\mathrm{p}>0.35$ is in the region
\begin{equation}
\eta > 3\times10^{-5}.
\end{equation}
Thus $\eta_{\rm b}=3\times10^{-5}$. For $\xi_e=-0.01$ and $\xi_e=0.1$, 
$\eta_{\rm b}=2\times10^{-5}$ and $\eta_{\rm b}=2\times10^{-4}$, respectively.
$\eta$ above these values are required to explain the bMS stars. 

\subsection{Upper Limits}
As previously predicted by Matsuura et al. \cite{matsu1,matsu2,matsu3},
heavy elements are produced during BBN with high $\eta$.
The amount of heavy elements produced during BBN gives the upper limit of $\eta$. Too large $\eta$ might overproduce  heavy elements incompatible with the metallicity of bMS, which is at least a factor of a few smaller than the solar abundances.
As nuclei with the mass number ($A$) of $\sim 80$ or greater are important for this
restriction, we use a large reaction network for this purpose.
For calculations of the large reaction network, we only consider the
case of $\xi_e=0$.

\begin{figure}[t]
\begin{center}
\includegraphics[height=8cm,width=9cm,clip]{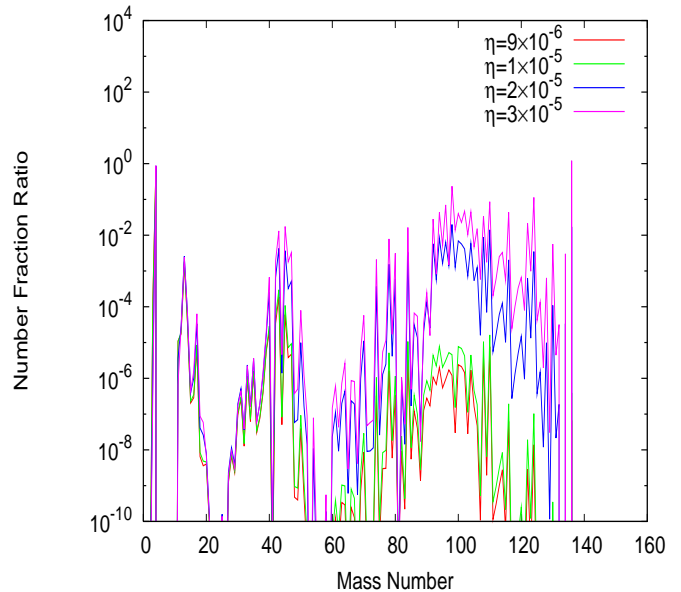}
\caption{The number fractions of the produced elements relative to
the solar abundances. All the calculations are done with $\xi_e=0$.}
\label{900}
\end{center}
\end{figure}

In Fig.~\ref{900}, we show the results of BBN calculations for some values of $\eta$.
Between $\eta=1\times10^{-5}$ and $\eta=2\times10^{-5}$, there is a
considerable jump
in the amount of heavy elements with $A\sim 100$.
Fig. 7 of Matsuura et al. \cite{matsu2} shows that 
$\eta=10^{-4}$ yields a significant amount of heavy elements with $A\sim 150$, comparable
to the solar abundances. With higher $\eta$, the abundances of heavy elements continue
to increase.
Observations show that the amounts of
heavy elements are at least a factor of a few less than the solar
abundances in the relevant GC \cite{metal}.
Therefore the upper limit of $\eta$ is somewhere between $3\times10^{-5}$ and $10^{-4}$.

In Fig.~\ref{900}, some elements $A\sim 100$ are very close to the expected abundances of the GC stars. Thus, it is a good touchstone of our model
to check the abundances of the elements $A\sim 100$ in the dwarfs in bMS. 

\section{EVOLUTION OF HELIUM ENHANCED REGION}
Even if a primordial gas with high $Y_\mathrm{p}$ was once formed at the time of BBN, it
might be mixed with other gases in the outer regions and end up with a lower fraction of
helium. We briefly show that the effect of such mixing would not affect
the region with high helium abundance. Here, we do not only consider effects of
diffusion and accretion, but also take into account self-gravity as a competing process.

\subsection{Diffusion}
Helium ions in the region with a higher $Y_\mathrm{p}$ might diffuse outward the surrounding region with lower helium contents. 
  This effect takes place after the horizon scale exceeds the size of the region, that is, after the temperature becomes lower than $10^8-10^7$ K.
The diffusion equation for helium is written as
\begin{equation}
\frac{\partial n_{\mathrm{He}}}{\partial t}=D\nabla^2n_{\mathrm{He}},
\end{equation}
where $n_{\mathrm{He}}$ is the number density of helium and $D$ is the
diffusion coefficient of helium with electrons and its magnitude is estimated as
\begin{equation}
D=\frac{\lambda v}{3},
\end{equation}
where $\lambda$ is the mean free path of helium ions and
$v=\sqrt{{3k_\mathrm{B}T}/{4m_{\mathrm{H}}}}$ is the thermal speed of helium ions. Here the constant $m_\mathrm{H}$ denotes the mass of hydrogen.
The time scale $t_\mathrm{diff}$ of the diffusion is approximated as
\begin{eqnarray}
t_\mathrm{diff}&=&\Big|\frac{\partial \ln n_{\mathrm{He}} }{\partial t}\Big|^{-1}\nonumber \\
&=&\Big|D\frac{\nabla^2n_{\mathrm{He}}}{n_{\mathrm{He}}}\Big|^{-1}\nonumber \\
&\sim& \frac{R^2}{D},
\end{eqnarray}
where $R$ is the size of the region. 

\subsubsection{Before Recombination}
Before the recombination of electrons, photons and matters had been coupled and had the same temperature. The number density of helium during this epoch is proportional to the cube of the temperature and written as 
\begin{equation}
n_{\mathrm{He}}=\frac{Y_\mathrm{p}}{4}n_b=\frac{Y_\mathrm{p}}{4}\left(\frac{T}{T_0}\right)^3\eta
n_{\gamma
 0}.
\end{equation}
Here $T_0=2.73$ K is the present temperature of CMBR and $n_{\gamma
 0}$ is the present number density of photons and expressed as $n_{\gamma
 0}={2\zeta (3)}/{\pi^2}\left({k_\mathrm{B}T_0}/{\hbar
 c}\right)^3=416\ \mathrm{cm^{-3}}$ where $\hbar$ is the reduced Planck constant.
For the case of $\eta=10^{-4}$ and $Y_\mathrm{p}=0.4$, 
\begin{equation}
n_{\mathrm{He}}=2\times10^{-4}T^3\ \mathrm{cm^{-3}}.
\end{equation}
Thus, the size $R$ is expressed as a function of the temperature
\begin{eqnarray}
R&=&\left(\frac{3Y_\mathrm{p}10^6M_\odot}{16\pi m_\mathrm{H}n_\mathrm{He}}\right)^{1/3}
=\frac{5\times10^{21}}{T}\ \mathrm{cm}.\label{size}
\end{eqnarray}

Next, we estimate $D$ before the recombination.
Assuming that the universe has no net charge, the number density of electrons
$n_e$ is written as
\begin{eqnarray}
 n_e&=&n_\mathrm{H}+2n_\mathrm{He}\\
&=&\left(1-\frac{Y_\mathrm{p}}{2}\right)\left(\frac{T}{T_0}\right)^3\eta n_{\gamma0}\\
&=&2\times10^{-3}T^3\ \mathrm{cm^{-3}}.
\end{eqnarray}
The last value is, again, for the case of $\eta=10^{-4}$ and $Y_\mathrm{p}=0.4$.
From~\cite{diffco},
\begin{eqnarray}
\lambda&=&\frac{9k_\mathrm{B}^2T^2}{4n_e
 e^4\ln\Lambda}=\frac{8\times10^5T^2}{n_e\ln\Lambda}\ \mathrm{cm},
\end{eqnarray}
where
\begin{equation}
\Lambda=\frac{3}{4e^2}\left(\frac{k_B^3T^3}{\pi n_e}\right)^{1/2}
=7\times10^3\left(\frac{T^3}{n_e}\right)^{1/2}.
\end{equation}
Thus,
\begin{eqnarray}
D=\frac{\lambda v}{3}=\frac{8\times10^{11}}{T^{1/2}}\ \mathrm{cm^2\, sec^{-1}}.
\end{eqnarray}

From $R$ and $D$ derived above, $t_\mathrm{diff}$ is estimated  to be 
\begin{equation}
t_\mathrm{diff}\sim\frac{3\times10^{31}}{T^{3/2}}\ \mathrm{sec}\sim9\times10^{20}\ \left(\frac{T}{10^7\mathrm{ K}}\right)^{3/2}\mathrm{sec}.
\end{equation}
This diffusion time scale is so large that the effect of diffusion is negligible.

\subsubsection{After Recombination}
Matters are assumed to completely decouple from radiation after the recombination. Thus,
$n_\mathrm{He}$ and $R$ after the recombination are denoted as  functions of the gas temperature $T$, 
\begin{equation}
n_{\mathrm{He}}=\frac{Y_\mathrm{p}}{4}\eta n_{\gamma
 0}\frac{\left(T_\mathrm{re}T\right)^{3/2}}{T_0^{3}}=80T^{3/2}\ \mathrm{cm^{-3}},
\end{equation}
and
\begin{equation}
R=\frac{7\times10^{19}}{T^{1/2}}\ \mathrm{cm}.
\end{equation}
Here the temperature $T_\mathrm{re}$ when free electrons recombine with hydrogen ions is about 5,500 K due to the higher density. We have assumed that all of helium and hydrogen atoms are neutral in this stage. Therefore the cross section for helium atoms becomes of the order of $\sim10^{-16}\ \mathrm{cm^2}$. Accordingly, $D$ becomes
\begin{eqnarray}
D=\frac{3\times10^{17}}{T}\ \mathrm{cm^2\,sec^{-1}}.
\end{eqnarray}
The diffusion timescale is obtained as a constant value.
\begin{equation}
t_\mathrm{diff}=2\times10^{22}\ \mathrm{sec}.
\end{equation}
This diffusion time is also too large.

\subsection{Accretion}
Since the region with enhanced helium content has a much higher density than the surrounding region, the region accretes matters from the outside, which might dilute the
helium fraction. For simplicity, the timescale of this process is estimated assuming the Bondi
accretion \cite{Bondi_52}. The time scale of the Bondi accretion is inferred from the relation
\begin{equation}
\frac{\dot{M}}{M}=4\pi G^2M\frac{\rho(\infty)}{c_s^3(\infty)}\lambda(\gamma),
\end{equation}
where $M$ is the total mass of the region ($10^6M_\odot$ in this case),
$\rho(\infty)$ and $c_\mathrm{s}(\infty)$ are the density and the sound speed
at infinity. The eigen value of the steady state solution $\lambda(\gamma)$ is a function of the adiabatic index
$\gamma$ written as
\begin{eqnarray}
\lambda(\gamma)&=&\left\{ \begin{array}{ll}
\left(\frac{2}{5-3\gamma}\right)^{\frac{5-3\gamma}{2(\gamma-1)}}&
(\gamma\ne\frac{5}{3}) \\ \\
e^{5/3} & (\gamma=\frac{5}{3}). \\
\end{array} \right.
\end{eqnarray}
The values of $\rho(\infty)$ and $c_\mathrm{s}(\infty)$ are taken from those in the surrounding region with $Y_\mathrm{p}=0.25$ and
$\eta=6\times10^{-10}$. They change as the universe expands and can be written as functions of temperature,
\begin{eqnarray}
\rho(\infty)&=&\left\{ \begin{array}{ll}
m_\mathrm{H}\eta n_{\gamma 0} \left(\frac{T}{T_0}\right)^3 \\ \\
m_\mathrm{H}\eta n_{\gamma 0}\left(\frac{\sqrt{T_\mathrm{re}T}}{T_0}\right)^3\\
\end{array} \right.
\\
&\sim&\left\{ \begin{array}{ll}
2\times10^{-32}T^3\ \mathrm{g\,cm^{-3}}& (\mathrm{before\ recombination}) \\ \\
3\times10^{-27}T^{3/2}\ \mathrm{g\,cm^{-3}} & (\mathrm{after\ recombination}), \\
\end{array} \right.
\end{eqnarray}
where  $T$ after the recombination denotes the gas temperature. Similarly, $c_\mathrm{s}(\infty)$ is expressed as
\begin{eqnarray}
c_s(\infty)&=&\left\{ \begin{array}{ll}
\frac{c}{\sqrt{3\left(1+\frac{3c^2\rho(\infty)}{4a_\mathrm{B}T^4}\right)}}\\ \\
\sqrt{\gamma\frac{k_\mathrm{B}T}{\mu m_\mathrm{H}}} \\
\end{array} \right.\\
&\sim&\left\{ \begin{array}{ll}
\frac{2\times10^{10}}{\sqrt{1+\frac{3\times10^3}{T}}}\ \mathrm{cm\,sec^{-1}}& \mathrm{(before)}\\ \\
10^4\sqrt{T}\ \mathrm{cm\,sec^{-1}}& \mathrm{(after)},
\end{array} \right.
\end{eqnarray}
where $\mu$ is the mean molecular weight and $a_\mathrm{B}$ is the radiation energy constant
$a_\mathrm{B}={8\pi^5k_B^4}/{15h^3c^3}$. Thus, the accretion time scale $t_\mathrm{acc}$
becomes
\begin{eqnarray}\label{accretion}
t_\mathrm{acc}=\frac{M}{\dot{M}}=\left\{ \begin{array}{ll}
3\times10^{36}\left(1+\frac{3\times10^3}{T}\right)^{3/2}T^{-3}\ \mathrm{sec}&\mathrm{(before)} \\ \\
6\times10^{11}\ \mathrm{sec}&\mathrm{(after)}. \\
\end{array} \right.
\end{eqnarray}
This result means that the time scale of accretion before the recombination is
long enough to avoid dilution of helium, while the time scale after the recombination becomes significantly shorter than the Hubble time at that epoch. This indicates that  the average helium content reduces considerably. However, if we take the effect of self-gravity into account, star formation might precede the dilution of helium depending on the effectiveness of turbulent mixing. 

\subsection{Star Formation}
The Jeans length $\lambda_J$ of the system before the recombination is
\begin{equation}
\lambda_\mathrm{J}=\frac{c_\mathrm{s}}{\sqrt{G\rho}}=\frac{10^{27}T^{-3/2}}{\sqrt{1+\frac{3\times10^8}{T}}}\ \mathrm{cm},
\end{equation}
and the Jeans mass $M_J$ is
\begin{equation}
M_J=\frac{4}{3}\pi\lambda_J^3\rho=\frac{8\times10^{21}M_\odot}{\left(T+3\times10^8\right)^{3/2}}.
\end{equation}
Therefore as long as the matter is ionized the mass of the helium rich
region never exceeds the Jeans mass. On the other hand, the Jeans mass
in this region reduces to $2\times10^3\, M_\odot$ immediately after the
recombination. The helium rich region collapses on the time scale of
$\sim 1.5\times 10^{11}$ sec, much shorter than the accretion time scale
at the temperature of $T_\mathrm{re}=5,500$ K, {\it i.e.},
$3.5\times 10^{25}$ sec (see equation (\ref{accretion}) for before recombination).
This preceding collapse might avoid overall mixing of the helium rich matter with the accreted ambient matter. The resultant cloud consists of helium rich core surrounded by a primordial gas  with the normal helium content. 

Though there is no widely accepted scenario for the formation of a GC, we raise possible routes to realize GCs with multiple main sequence. Subsequent star formation both in the core and envelope could lead to a star cluster with double main sequence resembling $\omega$ Centauri.  To reproduce a GC like NGC 2808, two such clouds with different $Y_\mathrm{p}$'s in helium rich cores are needed to collide and trigger star formations in each of regions with three different $Y_\mathrm{p}$'s. The collision with a high velocity removes stars that have formed before. This may also reduce the star-to-star scatter in metallicities. As a result, stars in the thus formed cluster have main sequences with three different helium contents and have similar metallicities. These features are observed for stars in NGC 2808. Cloud-cloud collisions have been raised as a possible mechanism for the globular cluster formation \cite{gratton04}. Young stellar clusters found in merging galaxies \cite{trancho07} and near the center of the Milky Way galaxy \cite{stolte08} support this mechanism.

\section{CONCLUSIONS}
We calculated the BBN for several values of $\eta$ and $\xi_e$ and looked for the
parameter region where the helium enrichment in the GCs is reproduced.
We set lower limits of $\eta$ to be $\eta_{\rm b}=3\times 10^{-5}$ for $\xi_e=0$,
$\eta_{\rm b}=2\times 10^{-5}$ for $\xi_e=-0.01$, and $\eta_{\rm b}=2\times 10^{-4}$ for $\xi_e=0.1$
. The upper limits come from metal abundances and the
allowed parameter region for the case of $\xi_e=0$ is
$3\times10^{-5}<\eta<10^{-4}$. If an inhomogeneity caused by the Affleck-Dine
baryogenesis made a region with $\eta$ and $\xi_e$ in these criterions, the helium abundance is enhanced
and the region can end up with GCs, showing  multiple main
sequences. Our model predicts that the abundances of heavy elements with
$A\sim100$ is enhanced in the bMS stars.
Stars in our model are expected to have more concentrated  spatial distribution of the bMS stars than the rMS as was recently reported for $\omega$ Centauri \cite{omega3}
because the bMS stars are formed in the central part of the helium
enhanced region while the rMS stars are formed in the outer envelope in our model.

A problem of the Affleck-Dine baryogenesis is that it can be adopted to
antimatters as well as ordinary matters and can create regions with enhanced antimatters. As there are no observations of such regions with antimatters, there should be some
mechanism that stops the generation of antimatters or there must be another baryogenesis mechanism to exclusively enhance ordinary matters up to $\eta\sim3\times10^{-5}$ in small regions. 

\begin{acknowledgments}
We are grateful to an anonymous referee for helpful suggestions to improve the presentation. We wish to acknowledge K. Sato, N. Kohri, S. Matsuura, and K. Ichikawa for useful comments on numerical calculations of big bang nucleosynthesis. We also wish to thank M. Kawasaki and K. Kamada for discussions on baryogenesis and cosmology. This work has been partially supported by the Grants in Aid for Scientific Research (21018004) of the Ministry of Education, Science, Culture, and Sports in Japan.
\end{acknowledgments}

\end{document}